\documentclass[sigconf,authorversion]{acmart}

\usepackage{multirow}
\usepackage{xltabular}
\usepackage{array}
\usepackage{graphicx}
\usepackage{csquotes}
\usepackage{enumitem}
\usepackage[capitalise,nameinlink,noabbrev]{cleveref}
\newcounter{numquote}

\copyrightyear{2021}
\acmYear{2021}
\setcopyright{rightsretained}
\acmConference[ITiCSE '2021]{26th Annual Conference on Innovation and Technology in Computer Science Education V. 1}{June 26-July 1, 2021}{Virtual Event, Germany}
\acmBooktitle{26th Annual Conference on Innovation and Technology in Computer Science Education V. 1 (ITiCSE '2021), June 26-July 1, 2021, Virtual Event, Germany}\acmDOI{10.1145/3430665.3456304}
\acmISBN{978-1-4503-8214-4/21/06}

\newif\ifanonymous \makeatletter \if@ACM@anonymous \anonymoustrue \fi \makeatother
\newcommand{\anonymized}[2]{\ifanonymous\textit{#2}\else#1\fi}

\renewenvironment{quote}
  {\vspace{0.2em}\small\list{}{\rightmargin=1.5em \leftmargin=1.5em}\item\relax}
  {\vspace{0.1em}\endlist}
\newcommand{\q}[3]{\begin{quote}\textit{\enquote{#1}} [TA #2, #3]\end{quote}}

\begin{document}

\title[Challenges Faced by TAs in CS Education Across Europe]{Challenges Faced by Teaching Assistants\\ in Computer Science Education Across Europe}


\author{Emma Riese}
\affiliation{
  \institution{KTH Royal Institute of Technology}
  \country{Sweden}
}\email{riese@kth.se}
\orcid{0000-0002-4525-3568}

\author{Madeleine Lorås}
\affiliation{
  \institution{Norwegian University of Science and Technology}
  \country{Norway}
}
\email{madeleine.loras@ntnu.no}
\orcid{0000-0003-4152-775X}

\author{Martin Ukrop}
\affiliation{
  \institution{Masaryk University}
  \country{Czech Republic}
}
\email{mukrop@mail.muni.cz}
\orcid{0000-0001-8110-8926}

\author{Tomáš Effenberger}
\affiliation{
  \institution{Masaryk University}
  \country{Czech Republic}
}
\email{tomas.effenberger@mail.muni.cz}
\orcid{0000-0001-5601-926X}

\renewcommand{\shortauthors}{Riese et al.}

\begin{abstract}
Teaching assistants (TAs) are heavily used in computer science courses as a way to handle high enrollment and still being able to offer students individual tutoring and detailed assessments. TAs are themselves students who take on this additional role in parallel with their own studies at the same institution. Previous research has shown that being a TA can be challenging but has mainly been conducted on TAs from a single institution or within a single course. This paper offers a multi-institutional, multi-national perspective of challenges that TAs in computer science face. This has been done by conducting a thematic analysis of 180 reflective essays written by TAs from three institutions across Europe. The thematic analysis resulted in five main challenges: \textit{becoming a professional TA}, \textit{student focused challenges}, \textit{assessment}, \textit{defining and using best practice}, and \textit{threats to best practice}. In addition, these challenges were all identified within the essays from all three institutions, indicating that the identified challenges are not particularly context-dependent. Based on these findings, we also outline implications for educators involved in TA training and coordinators of computer science courses with TAs.
\end{abstract}



\maketitle

\section{Introduction}
\label{Sec:Intorduction}

To teach and to learn computer science (CS) has been viewed as challenging and difficult by many previous studies~\cite{guzdial_learner_2015}. Enrollment in CS courses at the university level has continued to increase~\cite{sax_examining_2017}. Specifically, at the introductory level, courses with hundreds or even thousands of students are not rare. To manage these courses, teaching assistants (TAs, students who are employed to assist the faculty), are commonly used in CS~\cite{mirza_systematic_2019}. However, the TA perspective is not fully explored, and previous research has mostly reported on experiences from a single institution, course, intervention, or TA training initiative~\cite{mirza_systematic_2019}. To the authors' knowledge, no previous multi-national, multi-institutional study on TAs in CS have been conducted. This study aims to fill that gap by presenting and comparing data from three institutions in three different European countries. This paper aims to explore which main challenges TAs in CS face in their work and investigate whether or not the challenges differ between the three institutions. We define a challenge as something that is directly or indirectly described as an issue or difficulty. This paper is focused on two research questions (RQs):

\begin{enumerate}[label=(RQ\arabic*)]
\item Which challenges do TAs in computer science face?
\item Are the identified challenges similar or different across institutions and countries?
\end{enumerate}

\noindent
By increasing our knowledge and understanding of what our TAs experience as challenging in CS courses, instructors can make more informed decisions regarding their course structures and TA training. By providing a multi-national perspective on the perceived challenges, we aim to provide a more generalizable and nuanced picture relevant to the CS education community.

\section{Related Research and Theory}
\label{Sec:Related Research and Theory}

TAs have been employed to assist faculty in many CS courses at multiple institutions~\cite{mirza_systematic_2019}. Using TAs makes it possible to offer students individual guidance and feedback, also in large classes  ~\cite{reges_effective_1988, mirza_systematic_2019, ren_office_2019}. The TAs' work tasks differ between universities and courses, but often include conducting tutorials, assisting students during programming labs, developing course material, and assessing homework or exams~\cite{mirza_systematic_2019}. Grading students’ work, referred to as summative assessment~\cite{taras_formativeSummative_2005,dixson_formativeSummative_2016}, is not a work task for all TAs. Some universities have strict rules stating that TAs are explicitly not allowed to grade students, which is only carried out by senior staff members or faculty members~\cite{vihavainen_massive_2013,Loras_LearningA_2020}. On the other hand, TAs can be responsible for tutoring and providing feedback to the students throughout the course, referred to as formative assessment~\cite{taras_formativeSummative_2005, dixson_formativeSummative_2016}. The assessment carried out during a course should also be clearly linked to the intended learning outcomes and learning activities, referred to as constructive alignment~\cite{biggs_constructive_1996}. At some institutions, the TAs take an active role in constructing learning activities and assessment tasks~\cite{riese_tutoring_2020, Alvardo_microclasses_2017, Ukop_poster_2020, biggers_collaborative_2009}, which entails that the TAs are also contributing to the course structure and content to some extent. Previous research has also shown that TAs who conduct assessments in a group setting achieve higher reliability~\cite{harrington_TAparties_2018}, compared to in a solo-setting.

TAs have been found to be a contributing factor for student success~\cite{dechenne_modeling_2015}. Students can also view their TA as their main teacher within a course~\cite{riese_student_2020}, that is, the person they have most interactions with and turn to for help. The fact that TAs are themselves also students has been argued to make the TAs more approachable than professors or senior lectures~\cite{decker_lookingGlass_2006,ren_office_2019}. Furthermore, the fact that the TAs were, often recently, enrolled in a similar course helps them relate to the students and foresee possible misconceptions~\cite{ren_office_2019}. However, previous research has also shown that some TAs view their students as their friends, which can make the TA role challenging and can cause conflicts of interest to arise when grading~\cite{riese_tutoring_2020}. Both students~\cite{riese_student_2020} and TAs~\cite{riese_wip_2018, Muzaka_niche_2009, dechenne_modeling_2015} have experienced that they are not always properly trained for the TA role and lack pedagogical skills and pedagogical content knowledge (PCK). PCK, as introduced by Shulman~\cite{shulman_those_1986, shulman_knowledge_1987} is described as a combination or overlap between content knowledge and pedagogical knowledge, that is, knowledge on how to teach the specific content. This framework was later extended by Mishra and Koehler~\cite{mishra_technological_2006} to also include a technology knowledge component. The technology knowledge dimension intersects with both the pedagogical knowledge and the content knowledge dimension, in what is referred to as technological pedagogical content knowledge (TPACK)~\cite{mishra_technological_2006}. Both the PCK and the TPACK framework have previously been applied to CS contexts, mainly in the K-12 teacher education settings~\cite{vivian_identifying_2019,buchholz_pck_2013,brandes_using_2019}.

Efforts to support the TAs by offering TA training have been reported and presented in a number of publications, such as ~\cite{forbes_scaling_2017,brent_preparingUT_2007, estrada_bridging_2017,Ukrop_diary_2019, Loras_LearningA_2020, Ukop_poster_2020}. Furthermore, training has been reported to be an important factor in the TAs' professional development~\cite{marbouti_help_2013, moon_hinderhelp_2013}. One institution reported positive results with a team-teaching approach, where novice and experienced TAs were paired to work together when conducting tutorials~\cite{patitsas_case_2013}. The offering of introductory TA training has been suggested to bridge the gap between desired and actual competency among newly employed TAs~\cite{estrada_bridging_2017}. The social environment and the intensity of lab sessions have also been found to affect job satisfaction among TAs~\cite{patitsas_environmental_2012}. Some institutions have reported on high interest among students to become TAs~\cite{roberts_update_stanford_1995,vihavainen_massive_2013}, however, this is not the case at all institutions. How the TAs are recruited to the courses could also differ between universities~\cite{mirza_systematic_2019}, and a rubric to make the decision transparent and fair has been proposed in a previous study~\cite{leyzberg_nailing_2017}.

\section{Method} \label{Sec:Method}

\begin{table*}[t]
\caption{Comparison of the three participating institutions, their CS departments and TA situation}
\label{tab:research-setting}
\renewcommand\tabularxcolumn[1]{m{#1}}
\renewcommand{\arraystretch}{1.3}
\begin{tabularx}{\linewidth}{@{}l>{\raggedright\sc\arraybackslash}m{2.1cm}XXX@{}}
& & \textbf{\anonymized{NTNU, Norway}{Institution A}}& \textbf{\anonymized{KTH, Sweden}{Institution B}} & \textbf{\anonymized{MUNI, Czechia}{Institution C}} \\ \midrule
\multirow{4}{*}{\rotatebox{90}{\textsc{Institution}}} 
& Overall
    & research-focused university, 8 faculties, 7\,000 employees, 42\,000 students
    & research-focused university, 1 faculty, 5\,000 employees, 15\,000 students
    & research-focused university, 9 faculties, 6\,000 employees, 35\,000 students \\
& CS department
    & 3\,000 students, approx.\ 500 TAs
    & 4\,400 students, approx.\ 150 TAs
    & 2\,000 students, approx.\ 150 TAs \\
& CS courses
    & 7.5 ECTS, 50--3\,600 students/course
    & 3--9 ECTS, 20--250 students/course
    & 2--8 ECTS, 10--700 students/course \\
& TAs' level
    & bachelor, master, doctoral students
    & bachelor, master, doctoral students
    & bachelor, master, doctoral students \\ \midrule
\multirow{5}{*}{\rotatebox{90}{\textbf{TA position}}} 
& Responsibilities
    & hold open lab hours, assess assignments (often oral), facilitate project work, (rarely) lecture
    & hold open lab hours, assess assignments (often oral), grade exams, conduct tutorials
    & conduct tutorials, hold open lab hours, grade assignments, grade exams, (rarely) lecture \\
& Assessment 
    & cannot formally grade, but can assign pass/fail to assignments 
    & grade assignments (pass/fail or A--F), the examiner is formally responsible
    & grade assignments (pass/fail or points), sometimes grade exams \\
& Payment 
    & both teaching and preparation 
    & both teaching and preparation
    & both teaching and preparation \\
& Recruitment 
    & faculty-wide system based on grades and experience
    & course coordinators recruit independently based on their requirements
    & course lecturers recruit independently based on their requirements \\ \midrule
\multirow{3}{*}{\rotatebox{90}{\textbf{TA training}}}
& Format 
    & 20 hours; several teaching blocks throughout the semester
    & 6 hours; 3 online modules and 2 workshops before the semester
    & 30 hours; weekly seminars during the whole semester \\
& Participation 
    & only new TAs, mandatory
    & only new TAs, mandatory
    & any TAs, optional \\
& Compensation 
    & paid for the time in training 
    & paid for the time in training 
    & 3 ECTS credits for training \\
\end{tabularx}
\renewcommand{\arraystretch}{1}
\end{table*}

In order to investigate which challenges TAs in CS face (RQ1), we collected reflection essays from TAs from three different institutions in three different European countries (RQ2). This paper does not aim to evaluate the three institutions' use of TAs, but understanding their characteristics is important in order to understand the results. The different institutions and their TA programs are therefore described in \cref{tab:research-setting}. At each institution, we asked TAs to reflect on their own practice by answering these questions:

\begin{quote}\textit{\enquote{Describe an interesting situation or interaction you have experienced as an assistant. It can be something you found challenging, an ethical dilemma, or just something that has been on your mind. Reflect on how you handled the situation. What did you do well? What would you have done differently? Is there something you would like feedback on or questions you have?}}\end{quote}

\noindent
At \anonymized{NTNU and KTH}{institution A and B}, the essays were collected as part of introductory TA training courses. At \anonymized{NTNU}{institution A}, the data was collected during 2018 and 2019, towards the ends of respective semesters, and at \anonymized{KTH}{intuition B}, during the beginning of fall semester 2020. The essays were not graded on the content, but the TAs had to hand them in to complete their TA training course. At \anonymized{KTH}{institution B}, the TAs were also allowed to describe a fictive situation that they thought could occur since some of the TAs enrolled in the training course were very recently employed and had not yet gained much TA experience (but all of whom had been on the student side of TA-student interactions for years). At \anonymized{MUNI}{institution C}, the essays were collected during summer 2020 through the distribution of a digital survey asking the above-presented questions. The survey was distributed to all TAs enrolled in a voluntary TA training course in the previous four years and all the TAs of the second-largest undergraduate programming course. It was completely voluntary for all TAs to let their (anonymized) essays be part of this study, and informed consent was collected from all TAs. The data collection consist of 180 essays (119 from \anonymized{NTNU}{institution A}, 32 from \anonymized{KTH}{institution B} and 29 from \anonymized{MUNI}{institution C}). The essays were each half a page to a page long. A majority of the essays were written in the official languages of the given country, and a few were written in English. The essays from \anonymized{NTNU}{institution A} and \anonymized{KTH}{institution B} were analyzed in their original languages, while the essays from \anonymized{MUNI}{institution C} were first translated to English.

The essays were analyzed using a thematic analysis~\cite{braun_using_2006} aiming to identify common themes (the challenges) the TAs had written about. We followed the six steps outlined in~\cite{braun_using_2006}, but with some adaption to the specific data set at hand. The analysis was carried out for the data from one institution at the time and then merged at the final stage of the analysis. For the set of essays from each institution, two researchers first coded all essays independently and summarized both the initial codes, and identified themes of their respective analysis. The two researchers then met to discuss and compare the findings of their independent analysis. This resulted in an agreement of the final themes and codes identified for each subset of data. The analysis was also conducted with some time between to minimize the interference of the previously found themes. Once the analysis of all three subsets was completed, we started to view the data as a complete set and merged the identified codes and themes. While doing so, we created a copy of the codes and themes that omitted which institution they originated from. This was done to not be influenced by the origin of the codes (since RQ2 aims to investigate potential differences). This data was, however, kept separate so we cold backtrack and validate the origin after this step was completed, and the writing up of the results began. The final merging and formulation of the codes and themes was also carried out by both researchers independently, followed by a discussion resulting in the final themes. When this was completed, we revisited the essays and previous codes to validate our findings and backtrack the origin of themes. We also decided to cut out the parts of the data that were only about constraints or challenges caused by the COVID-19 pandemic since the data from \anonymized{NTNU}{institution A} was collected before the pandemic broke out.

\section{Results}
\label{Sec:Results}

\begin{figure*}[t]
    \centering
    \includegraphics[width=13cm]{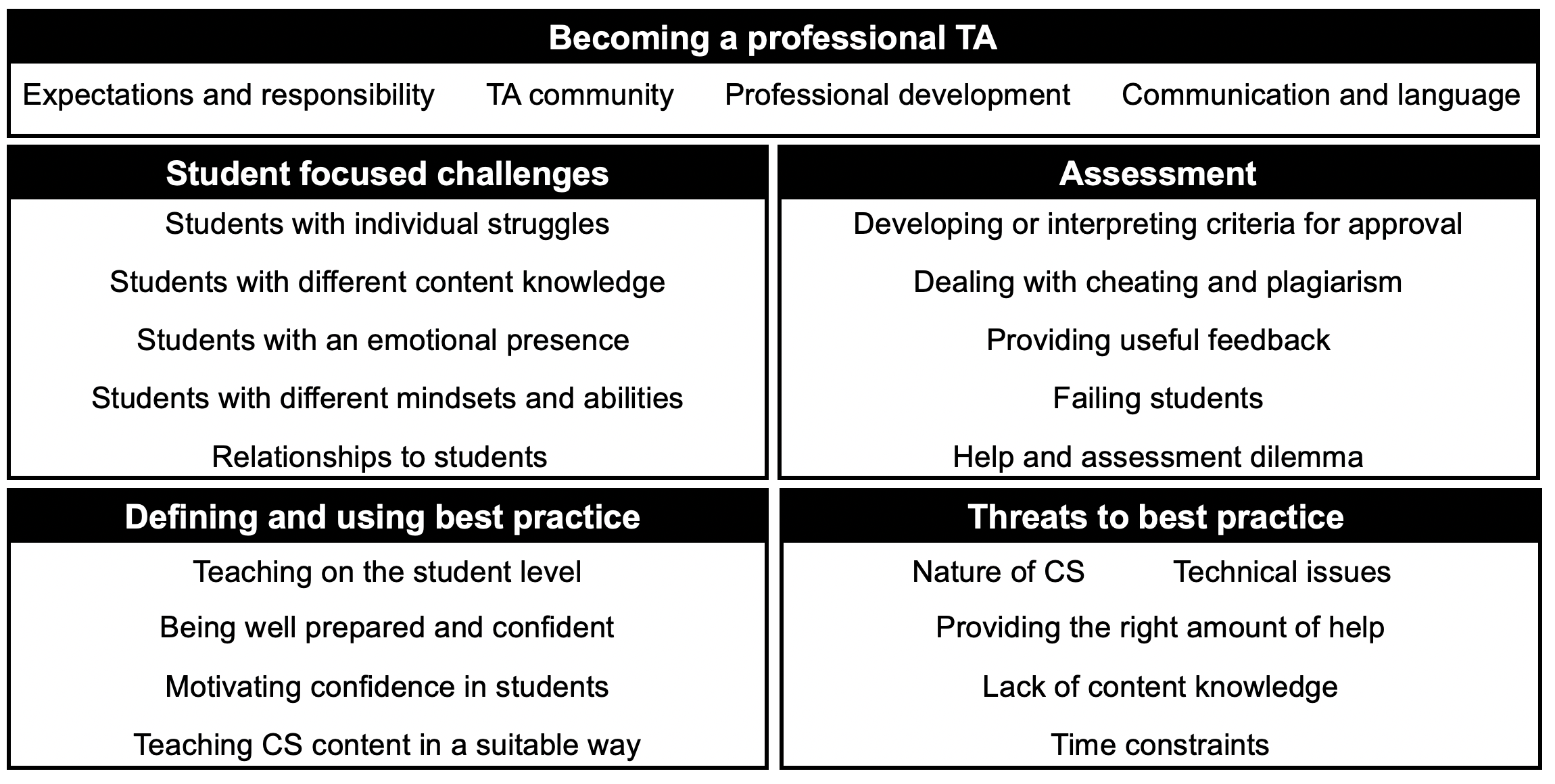}
    \caption{Main challenges identified in the thematic analysis}
    \label{fig:themes}
\end{figure*}

From the thematic analysis, we identified results along two central dimensions corresponding to the posed research questions: \textit{main challenges} and \textit{institutional similarities and differences}. The main challenges that were identified from the analysis are presented in \cref{fig:themes} along with all sub-themes corresponding to each main challenge.

\subsection{Main Challenges}
Each main challenge represents a theme describing an area TAs found challenging. Each theme consists of a set of sub-themes, aggregated from the codes and represent recurring topics in the TAs reflections across the three included institutions.\footnote{A complete overview of themes, codes, and exemplary quotes \anonymized{can be found at \href{https://doi.org/10.18710/O8FCIK}{https://doi.org/10.18710/O8FCIK}}{will be available online after the review}.} Although the main challenges are separate issues, TAs would often mention several main challenges in one example, illustrating that there are also complex interconnections present between the identified themes.

\subsubsection{Becoming a professional TA}

Being a professional was a topic many TAs reflected on as important but challenging. They described a mismatch of expectations between course teachers and students, especially in combination with unclear instructions. For example, course teachers would intend TAs to mainly facilitate group work, while the students expected debugging help and technical support. The TA community was seen as an important support network, but discovering unethical behavior of other TAs were sometimes mentioned as well. Multiple TAs noted areas they once found challenging but have since overcome, such as public speaking and personal interactions.

\subsubsection{Student focused challenges}

TAs reported on challenging experiences handling the vast diversity of students. From students with special needs, students dealing with personal problems to those unhappy about their assessment, or students working in groups. The level of content knowledge could also differ, from students who were very advanced to weaker students who were behind in the course. It was also described as challenging to meet unmotivated, passive, and unprepared students, as well as students who did not want help or were only focused on end results. The TAs also reflected on the uncomfortable interaction in situations with overly emotional, stressed, or upset students.

Multiple TAs reported that their own relationships with particular students sometimes made the interactions even more complicated. These included pre-existing friendly and romantic relationships, as well as the development of the TA-student relationship over the semester. Although many TAs highlighted the pedagogic benefits of having or developing strong relationships with their students, they also found it challenging to set boundaries and be professional, as is illustrated by the quote below.

\q{It is not always so easy to grade their [close friends'] papers and I feel a conflict of interest while doing so. [...] It does not always feel fair that I spend so much time guiding my friends and not the rest of the students. It seems that the threshold for asking a friend through a FB-message is much lower than sending an email to an unknown teaching assistant.}{48}{\anonymized{NTNU}{Institution A}}

\subsubsection{Assessment}

TAs found it challenging to develop and/or interpret passing/grading criteria, specifically mentioning determining what is \enquote{good enough}, treating all the students fairly, and assessing effort and group work. Not surprisingly, failing a student was considered challenging: TAs reported feeling pressure from the students and uncomfortable announcing that the student failed, particularly in face-to-face interactions within hearing distance of other students. A further challenging aspect was cheating. Both determining if the submitted work was, in fact, plagiarized and acting on suspicions of unethical behavior were described as common challenges. In the formative paradigm, TAs reported that providing useful feedback was challenging, especially to students who were solely focused on the end result. Furthermore, TAs often described the challenge of giving both formative and summative feedback to the same students, sometimes even in the same session (guiding a struggling student and then immediately assessing if the submission was adequate).

\subsubsection{Defining and using best practice}

Overall, many TAs found it challenging to identify good practice for efficient teaching and learning CS and in applying it in practice. The topics mentioned included, for instance, visualizing code flow, writing pseudo-code, being creative, problem decomposition, and planning before coding. As a prominent sub-theme, the TAs found it challenging to teach at the students' level of understanding. First, the TAs needed to map the students' knowledge and understanding of the topic and then try to support them from there. Specifically, it was challenging for the TAs to formulate good questions and work with the student. Helping students reflect on their work and supporting good study habits was also brought up as wanted best practices but experienced as challenging to put into practice. TAs also mentioned to struggle with how to properly prepare for tutorials and lab sessions, feeling confident in teaching, making students feel confident and motivating them, and handling arising conflicts.

\subsubsection{Threats to best practice}
\label{Sec:Threats}

Although TAs reported on a plethora of effective pedagogical and didactic strategies to help students learn, they often went hand in hand with a challenge. For example, giving feedback to students' code, debugging code, and using pair programming. Several TAs reflected on the fact that CS was a new and difficult topic for many students, especially in the introductory courses. The following quote illustrated another aspect tied to the CS content: There are often multiple ways to solve an assignment, making it difficult to assess.

\q{As a TA I have met students who solve lab assignments in a very different way than the course teacher's solution. [...] It is a lot harder for a TA to assess these kinds of solutions. First the TA must interpret what the goal of the assignment was and what the central aspects in functionality and interaction were.}{19}{\anonymized{KTH}{Institution B}}

\noindent
In project-based courses, the programming language and technology used is sometimes up to the students to decide, resulting in TAs having to support topics outside their expertise. Furthermore, TAs also reported technical issues that stand in the way of learning the content (e.g., IDEs, operating system, version control).

Furthermore, TAs described being insecure about their lack of content knowledge, especially with new material, and, in general, just being worried about giving out the wrong information to their students. Providing the right amount of help was a commonly mentioned challenge as well. Concretely, resisting the urge to take over the student's keyboard, not pushing your own solutions, and balancing help, guidance, and teaching.

The threat to adopting best practice that most TAs reported on was, however, in-class time constraints. The time challenge involved dividing time evenly, prioritizing students who needed help, assessing students who like to present their solutions, and giving time to advanced students.
The time predicament was visible throughout many identified themes, as illustrated by the quote below.

\q{In stressful situations, it can be easy to forget to take it slow to make sure the student understands fully what you are trying to help with. [...] In addition, I have a tendency to take over the students' keyboard when I feel the time pressure. One should always take the time to make sure the student has understood the problem you have been helping them with and adding some constructive feedback.}{6}{\anonymized{NTNU}{Institution A}}

\subsection{Institutional Similarities and Differences}

In order to address the second research question, we need to examine the similarities and differences across institutions and countries. As described in \cref{tab:research-setting}, the way TAs work differs somewhat at \anonymized{MUNI}{institution C} compared to the other two institutions. At \anonymized{NTNU and KTH}{institutions A and B}, TAs mostly help and support the students in open labs and determine if assignments are passed or failed. At \anonymized{MUNI}{institution C}, however, TAs have a more formal role, mostly conducting tutorials, assigning homework, and grading. Additionally, the TAs at \anonymized{KTH and MUNI}{institutions B and C} plan and conduct tutorials, while TAs at \anonymized{NTNU}{institutions A} have less responsibility in the planning and are just there to answer questions and conduct pass/fail assessment. 

However, neither of these differences was prevalent in the main challenges presented in the previous section. While we found that the different structures lead to different specific situations, we also found that the core challenges remained the same across the studied institutions. TAs at \anonymized{NTNU and KTH}{institutions A and B} would describe time management issues with students one-to-one or in the queue, while TAs at \anonymized{MUNI}{institution C} would discuss how to divide the time during a tutorial. An example becomes apparent with this reflection from a TA at \anonymized{MUNI}{institution C}: 

\q{I tried from the beginning to explain the most important things, so that most of them [students] at least had a chance to \enquote{catch the train}, but it was at the expense of the time spent working on exercises.}{23}{\anonymized{MUNI}{institution C}}

\noindent
Comparing the statement above to the second quote from \anonymized{NTNU}{institution A} in \cref{Sec:Threats} about time management, it is evident that even though the specific situations were different, the core challenge was time constraints. At \anonymized{NTNU}{institution A}, the challenge was how to manage the time when helping students individually and to use best practices under stress. At \anonymized{MUNI}{institution C}, the challenge was how to divide the time in a tutorial between revising information from the lecture and working on exercises. These are both examples of the time constraints theme, but with different specific situations in different educational structures at the two universities. 

Similarly, TAs at \anonymized{NTNU and KTH}{institutions A and B} who did not formally decide on the grading experienced similar challenges regarding assessment as the TAs at \anonymized{MUNI}{institution C} who have that responsibility. Passing/failing assessments were described as similarly challenging, regarding assessing friends, setting the standard, and giving feedback, as actually setting a grade. Therefore, it can be concluded that all main challenges described were found to be similar across the examined institutions and countries, regardless of the education structure. 

\section{Discussion}
\label{Sec:Discussion}

We have identified five main challenges that the TAs at \anonymized{NTNU, KTH and MUNI}{institutions A, B and C} face: \textit{becoming a professional TA}, \textit{student focused challenges},\textit{assessment}, \textit{defining and using best practice}, and \textit{threats to best practice}. In many regards, these results confirm previous findings about the TAs' experiences. To begin with, our results strengthen the claim that TAs need help and support to develop within their role, which has also previously been shown~\cite{patitsas_case_2013, marbouti_help_2013}. The TA community and social environment were shown to play a key role in that, which is also aligns with previous findings~\cite{patitsas_environmental_2012}. Communicating with the students and tutoring them is also a big part of the TAs' work tasks~\cite{marbouti_help_2013, riese_wip_2018} and our analysis found that this also comes with a whole set of challenges. Conducting assessments have previously been reported as difficult for TAs~\cite{riese_wip_2018, marbouti_help_2013, harrington_TAparties_2018}. Furthermore, it has been found that TAs are both the tutor and grader to the same students~\cite{riese_tutoring_2020}, two roles that are non-trivial to combine. 

The fact that TAs experience that they are approachable to their students~\cite{ren_office_2019}, also comes with the downside of being too close to their students. A previous study reported that the TAs could view themselves as friends to their students~\cite{riese_tutoring_2020}, and our results extend on that. Our results show that personal relationships between TAs and their students exist and could be challenging for the TAs to handle. Time constraints have also been found to be hindering and challenging in previous studies~\cite{marbouti_help_2013,riese_wip_2018}, which was confirmed by this study. 
The major challenges that have not received the same focus in previous studies are defining and using best practice. Although there have been studies reporting on training initiatives for TAs in CS~\cite{forbes_scaling_2017,brent_preparingUT_2007, estrada_bridging_2017,Ukrop_diary_2019, Loras_LearningA_2020, Ukop_poster_2020}, little emphasis has been put on the CS specific best practices that we found the TAs also face. It is not surprising since CS is considered hard to both teach and learn~\cite{guzdial_learner_2015}. Our findings, point towards challenges that span through the whole TPACK (technological, pedagogical and content knowledge) framework~\cite{mishra_technological_2006}, including content knowledge, pedagogical knowledge, technology knowledge, and where they intersect. It is noteworthy, that TAs who have all been studying CS themselves experienced technical issues, and found the use of new software challenging. This has not been reported before and shows that it can not be taken for granted that you are an expert on all new technology simply because you are a CS TA.

The previous studies to which we have compared our results have, however, been conducted in small scales, isolated to one institution or one course. With our findings regarding RQ2, we could see that the identified challenges are present at multiple institutions in multiple countries. It is also interesting to note that these challenges were found to be similar across the institutions, despite different organizational structures. We would like to emphasize that TAs who take an active role in assigning homework or designing course material need to be aware of the intended learning outcomes, in order to be able to achieve constructive alignment~\cite{biggs_constructive_1996} in the courses. The same applies for TAs who are tasked with using and interpreting passing/grading criteria. In order for the TAs to be successful and follow the course requirements, they need to understand the aim of each assignment they grade and each tutorial they conduct.

\subsection{Implications}
\label{Subsec:Implications}
Based on these findings, we would like to highlight some implications for educators involved in TA training and coordinators of CS courses with TAs. The presented recommendations can also work as pointers for future research studies since whether or not they do have a positive effect on the TA experience remains to be investigated. Since this study shows that the identified challenges are similar between the three studied institutions, we would also argue that smaller scaled studies to address these can be valuable for the CS education community.

\subsubsection{Be aware that best practice needs to be defined and spread.} We can not assume that TAs have all the necessary content, pedagogical and technological knowledge needed as soon as they start. However, we can help TAs define best practices in CS education and share examples of how that can be implemented. Facilitating a social and supportive community is also believed to aid the sharing of best practices among TAs. As shown, the TA community can sometimes lead to the reproduction of unethical or unproductive practices, and we would therefore argue that experience is not enough. Formal TA training that includes illustrative examples is also needed. 

\subsubsection{Acknowledge threats to best practice and address them.} Being aware of the main threats to best practice could be seen as a first step in overcoming these. Course coordinators have the power to make informed decisions to minimize these threats. For instance, one of the identified threats was working under time constraints. If you expect your TAs to be able to give the students detailed feedback and carefully guide them through a difficult programming problem, the TAs need to have sufficient time to do so. Another example is that material and instructions for assessment need to be clear, and even if they are, the TAs might still need additional help interpreting and using them. 

\subsubsection{Dare to discuss ethical dilemmas and provide guidelines.} Dealing with ethical dilemmas, such as suspicions of plagiarism or deadline extensions for desperate students, is something that needs to be addressed and discussed with the TAs. The TAs need to know how to handle such situations and take actions based on knowledge, not feelings. Even though this might seem trivial to an experienced course coordinator, it is not trivial to all TAs.

\subsubsection{Recognize the student-TA relationship as unique.} The social aspect of being a TA and the interaction with students are crucial parts of the TAs' work. Concurrently, the student-TA relation was experienced as a major challenge for many TAs in this study. The TAs need to be equipped with tools and techniques to be able to overcome these challenges sufficiently. Some of the described challenges were related to general pedagogic knowledge, such as motivating students and handling a diverse student group. Other challenges come from the fact that TAs have other types of relationships with their students (for instance, are friends with or even romantically involved with their students). These are believed to be specific to TAs and need to be addressed as such. To the faculty that train TAs, we would recommend addressing these and providing the TAs with a good foundation on how to handle specific situations.

\subsection{Threats to Validity and Limitations}
\label{Subsec:Threats to Validity and Limitations}
In this research, we have used a qualitative method, with a large sample size (180 participants). Nevertheless, it is important to note that the sample size between the three institutions differed, which could have had an impact when comparing the three data sets to each other. If a theme was not present in the data set for an institution, that does not necessarily imply that the TAs have not experienced that challenge since we did not ask the TAs to name all challenges they ever encountered. However, we did not find any major differences between the institutions. This finding both strengthens the claim that the identified themes were truly the main challenges across the institutions and that even the smaller sample sizes (29–32) were sufficient to capture these through asking open-ended questions. It should also be noted that the open-ended questions did not explicitly ask the TAs to name their main challenges but rather to describe a situation or interaction and reflect upon it. It is, of course, a possibility that the TAs would have written something else if asked explicitly, but this method was chosen to give us the teaching contexts and enable the TAs to describe a challenging situation without having to pinpoint a specific challenge. The data is also limited by what the TAs were comfortable sharing. All collected data are also self-reported by the TAs – asking the students and course coordinators or lecturers for their view on these challenges would be an interesting additional input and possibility to validate the results further.

A limitation of the setup of this study is that we only studied three institutions within Europe, and the generalizability to other institutions is not investigated. In this research, we did also not take into account how much experience the TAs have had prior to writing their reflections, which could impact what they wrote in their essays. The data were not collected with prior experience as a controlled variable. At \anonymized{KTH}{institutions B}, a majority of the TAs were new TAs, writing these essays at the beginning of the semester, at \anonymized{NTNU}{institutions A} the TAs wrote these essays at the end of their first semester, and at \anonymized{MUNI}{institutions C} it was more scattered. The unified results do suggest that the identified challenges are found across institutions and among TAs with different long experience, but we can not make any claims on to which degree experience played a role from this study. The presented results should also not be seen as a complete list of challenges that TAs in CS face. That was also not the aim of this research, and the results should rather be seen as a list of main challenges identified across the three studied institutions. In this study, all steps in the thematic analysis were carried out by two researchers independently, followed by a discussion resulting in an agreement. This rigorous process strengthens the trustworthiness of the results.

\section{Conclusions}
\label{Sec:Conclusions}
In this study, we have identified five main challenges that TAs in CS face by analyzing 180 reflective essays from TAs from three different institutions in three different countries. We also found that the identified challenges (\textit{becoming a professional TA}, \textit{student focused challenges}, \textit{assessment}, \textit{defining and using best practice}, and \textit{threats to best practice}) were present in the essays from all three intuitions. In fact, no major differences were found between the institutions, despite the different organizational setups. We conclude by emphasizing that TA training and support are needed in order to assist the TAs in overcoming these challenges.



\bibliographystyle{ACM-Reference-Format}
\bibliography{main}

\end{document}